\documentclass[12pt]{article}
\pdfoutput=1
\usepackage{cite}
\usepackage{amssymb}
\usepackage{amsmath}
\usepackage{bm} 
\usepackage{graphicx}
\usepackage{color}
\usepackage{xcolor}
\usepackage{dsfont}
\usepackage{enumerate}
\usepackage{hyperref}
\usepackage{setspace}
\usepackage[latin1]{inputenc}
\definecolor{nicered}{rgb}{0.7,0.1,0.1}
\definecolor{nicegreen}{rgb}{0.1,0.5,0.1}
\hypersetup{colorlinks,citecolor= nicegreen,linkcolor= nicered}

\newcommand{\be}  {\begin{equation}}
\newcommand{\ee}  {\end{equation}}

\def\e6{E(6)}
\def\10{SO(10)}
\def\21{SA(2) $\otimes$ U(1) }
\def\321{$\mathrm{SU(3) \otimes SU(2) \otimes U(1)}$ }

\def\422{SA(4) $\otimes$ SA(2) $\otimes$ SA(2)}
\def\roughly#1{\mathrel{\raise.3ex\hbox{$#1$\kern-.75em
      \lower1ex\hbox{$\sim$}}}} \def\lsim{\roughly<}
\def\gsim{\roughly>}

\def\lsim{\raise0.3ex\hbox{$\;<$\kern-0.75em\raise-1.1ex\hbox{$\sim\;$}}}
\def\gsim{\raise0.3ex\hbox{$\;>$\kern-0.75em\raise-1.1ex\hbox{$\sim\;$}}}
\nopagebreak
\allowdisplaybreaks
\begin{document}
\begin{titlepage}


  \newcommand{\AddrIPM}{{\sl \small School of physics, Institute for
      Research in Fundamental Sciences (IPM),\\ \sl \small
      P.O. Box 19395-5531, Tehran, Iran}}
        \newcommand{\AddrDenmark}{{\sl \small Department of Physics and Astronomy, Aarhus University,\\ \sl \small
    8000 Aarhus C, Debmark}}
  \vspace*{0.5cm}
\begin{center}
  \textbf{\large Neutrinos secretly converting to lighter particles to please both KATRIN  and the cosmos }

  $^a$Yasaman Farzan\footnote{e-mail address:{\tt yasaman@theory.ipm.ac.ir}} and $^b$Steen Hannestad\footnote{e-mail address:{\tt sth@phys.au.dk}}
  \vspace*{0.4cm}\\
  $^a$\AddrIPM.
    \vspace*{0.4cm}\\
  $^b$\AddrDenmark.
  \vspace{1cm}\\
\end{center}
\vspace*{0.2cm}
\begin{abstract}
 \onehalfspacing
Within the framework of the Standard Model of particle physics and standard cosmology, observations of the Cosmic Microwave Background (CMB) and Baryon Acoustic Oscillations (BAO) set stringent bounds on the sum of the masses of neutrinos. If these bounds are satisfied, the upcoming KATRIN experiment which is designed to probe neutrino mass down to $\sim 0.2$~eV will observe only a null signal.
We show that the bounds can be relaxed by introducing new interactions for the massive active neutrinos,  making neutrino masses in the range observable by KATRIN compatible with cosmological bounds. Within this scenario, neutrinos convert to new stable light particles by resonant production of intermediate states around a temperature of $T\sim$ keV in the early Universe, leading to a much less pronounced suppression of density fluctuations compared to the standard model.
\end{abstract}
\end{titlepage}
\setcounter{footnote}{0}
\section{Introduction
}\label{sec:intro}
In recent years various solar, atmospheric, long baseline and reactor neutrino experiments have shown that the flavor of neutrino beams traveling over relatively large macroscopic distances can change. Neutrino oscillation within the three neutrino scheme is given by two mass square splittings ($\Delta  m_{21}^2$ and $\Delta  m_{31}^2$), three mixing angles ($\theta_{12}$, $\theta_{23}$ and $\theta_{13}$) and a CP-violating phase ($\delta_D$). All these parameters, except for $\delta_D$ and ${\rm sign} (\Delta m_{31}^2)$, have already been measured with a remarkable  precision (see e.g.\ \cite{Gonzalez-Garcia:2014bfa} for a recent overview). However, the overall scale of neutrino mass or in other words, the mass of lightest neutrino is not yet known.

Information on the overall scale of neutrino mass can be obtained by measuring the distortion of the endpoint of the electron spectrum emitted in beta decay. The strongest bound so far was obtained by the Mainz experiment by studying the endpoint of the electron spectrum in Tritium decay ($^3{\rm H}\to ^3{\rm He} +\bar{\nu}_e+e$). The Mainz upper bound on neutrino mass is $2.2$~eV \cite{Mainz}.
KATRIN (KArlsruhe TRItium decay Neutrino) experiment \cite{Osipowicz:2001sq} is designed to probe $m_{\nu_e}$ down to 0.2 eV at 90 \% C.L. with a detection limit of 0.35~eV ($5\sigma$) \cite{0.35}
\footnote{KATRIN is scheduled to start taking data in 2016 \cite{Mertens}.}.

On the other hand, nonzero neutrino mass can dramatically affect cosmological structure formation by suppressing the growth of fluctuation on scales below the free-streaming scale.
A sum of neutrino masses saturating the bound from the Mainz experiment would have been easily visible in current Cosmic Microwave Background (CMB) and Large Scale Structure data.
In fact, data from the Planck satellite mission measurements of the CMB \cite{Ade:2015xua} provide an upper limit on the neutrino mass of
$\sum_i m_{\nu_i} < 0.71$ eV, already close to the projected sensitivity of KATRIN.
When auxiliary data from measurements of baryon acoustic oscillations (BAO)
is also used the bound is strengthened to
$\sum_i m_{\nu_i} < 0.23$ eV \cite{Ade:2015xua}.
Even a sum of neutrino masses as small as $\sum_i m_{\nu_i} \sim 0.06$ eV, the minimum allowed in the normal hierarchy, leads to a suppression in power of several percent, enough to be seen by future high precision surveys such as EUCLID \cite{Laureijs:2011gra,Hamann:2012fe,Basse:2013zua,Audren:2012vy}.

Thus, the cosmological bound on the sum of masses naively implies that  KATRIN will not be able to discern the effect of neutrino masses.
 A measurement of a non-zero neutrino mass by KATRIN will therefore have profound implications for cosmology and particle physics, making it imperative to reconsider the standard assumptions on cosmic evolution and neutrino properties that have been made to derive the cosmological bound on neutrino masses. In   this paper we propose two possible scenarios which make the relatively large neutrino masses measurable at KATRIN compatible with cosmological bounds by introducing new particles coupled to neutrinos. The scenarios are based on the following mechanism: After the Big Bang Nucleosynthesis (BBN) era and before recombination epoch (eV $<T<1$ MeV) the massive active neutrinos (partially) convert to light degrees of freedom either through coannihilation or through scattering off dark matter. As a result, the bounds from structure formation on the sum of neutrino masses can be relaxed, making neutrino masses heavy enough to be discerned by KATRIN cosmologically acceptable.

The new interaction has to be strong enough to efficiently convert neutrinos  to lighter new particles before $T\sim 1~e$V in the early universe. On the other hand, remaining neutrinos and the new particles should freely stream at the recombination era ($T\sim 0.3$ eV) \cite{secret}. Moreover, if the conversion of neutrinos to the lighter new particles takes place before neutrinos decouple from the standard model sector ($T>{\rm MeV}$), the new particles will contribute to extra relativistic degrees of freedom on which there are strong and relatively robust bounds from BBN and CMB (see e.g.\ \cite{Ade:2015xua}). Satisfying all these three conditions makes it challenging to come up with a consistent scenario.

In section \ref{general}, we discuss the general features of scenarios that convert neutrinos to light new particles during the epoch $1~{\rm eV}\ll T\ll 1~{\rm MeV}$.
We find that resonant scattering or coannihilation can be used to avoid too large effects for both $T>1$ MeV and $T<10$ eV but still achieve efficient conversion in the interval between these two epochs. In section \ref{model}, we will  present  low energy models within which such resonances can occur and discuss the bounds from various cosmological and astrophysical observations as well as terrestrial experiments on the parameters of the model. In section \ref{Cosmo}, we discuss how many new degrees of freedom are required to make cosmological bounds on sum of the masses of neutrinos compatible with relatively large neutrino mass measurable at KATRIN.
Our findings are summarized in section  \ref{Conclusions}.

\section{General features of the scenario \label{general}}
There are (at least) two possibilities to convert active neutrinos to lighter species at temperatures $T \lesssim m_e$: (1) new coannihilation modes of active neutrino pairs and (2) scattering of neutrinos off dark matter. In this section, we first briefly discuss the conversion mechanism for each case and then discuss the general effects of back reaction for both cases. In the end, we quantify the effective number of massive neutrinos after conversion.

Let us first discuss the case of neutrino pair coannihilation.
    If the mass of intermediate state responsible for coannihilation ($m_X$) is much larger than  the temperature, the coannihilation rate will be proportional to $T^5/m_X^4$ which should be compared to Hubble expansion rate $T^2/M_{Pl}^*$. Coannihilation would be therefore more efficient at higher temperatures when neutrinos were still in thermal equilibrium  and conversion to lighter new states would enhance the number of extra relativistic degrees of freedom on which there are strong bounds \cite{extraPLANCK}. On the other hand, at $T\gg m_X$, the coannihilation rate will be proportional to $T$. Comparing to the Hubble expansion rate then implies that the coannihilation becomes more efficient at lower temperatures so there would be no danger of producing extra relativistic degrees of freedom before neutrino decoupling. However, through the same interactions, neutrinos and new particles produced by coannihilation will scatter off each other with rate again given by $T$. Comparing to the  Hubble expansion rate $T^2/M_{Pl}^*$, we find that scattering becomes more important at lower temperatures.
As a result for $m_X<1$ eV, if the couplings are large enough for efficient neutrino conversion to new states, they cannot freely stream at the  time of recombination which is a requirement for successful structure formation (see e.g.\ \cite{secret}).

Thus, neither for $m_X>1$ MeV nor  for $m_X<1$ eV, the bounds can be satisfied. For eV $<m_X<1$ MeV, the $X$ particles can be resonantly produced and subsequently decay into new light states for $T\sim m_X$ converting a substantial fraction of active neutrinos to lighter states. In the appendix,  using narrow width approximation, we calculate the conversion rate ($R$) of a neutrino with a given momentum coannihilating  with any other neutrino in a medium. Using the formulas in the appendix it is straightforward to show that the coupling of resonant states to active neutrinos has to be larger than $5\times 10^{-11}(m_X/{\rm keV})^{1/2}$ to fulfill the requirement for efficient conversion; {\it i.e.,} $R\cdot H^{-1}>1$. Because of resonance enhancement, coupling so small will be enough to efficiently convert neutrinos at $T\sim m_X$. However, at $T\gg m_X$ or at $T \ll m_X$, this new coupling will be irrelevant  because (i) it cannot give rise to a significant deviation of $N_{eff}$ from 3 and (ii)  it cannot hinder the free streaming at recombination era. In the next section, we will present two models within which the resonant conversion scenario can be naturally embedded.

Let us now discuss scattering of active neutrinos off background Dark Matter (DM) particles  {\it i.e.,
} $\nu +{\rm DM}\to f +{\rm DM}$ where $f$ is the final particle which is even lighter than neutrinos. For DM mass larger than MeV, during epoch of our interest, DM particles are non-relativistic. We generally expect ({\it e.g.,} within thermal freeze-out scenario) that the number density of DM particles  has been fixed by $T\sim 1$~MeV. Considering that the average energy density of DM today is of order of $\rho_0\sim {\rm keV~ cm}^{-3}$, the number density of DM particles at $T<$MeV is given by  \be n_{DM}=\frac{\rho_0}{m_{DM}}\frac{ T^3}{T_0^3}. \label{nDM}\ee In general, we expect the scattering cross section ($\sigma_{scatt}$) to be proportional to $E_\nu^2/(m_{DM}^2-m_X^2)^2$. The scattering rate will  then be given by $n_{DM}\sigma_{scatt}\propto T^5/[m_{DM}(m_{DM}^2-m_X^2)^2]$. Comparing to $H\sim T^2/M_{Pl}^*$, we find that the scattering would be more efficient at higher temperatures when neutrinos have not decoupled so the scattering would contribute to extra relativistic degrees of freedom. However, if the splitting between $m_{DM}$ and $m_X$ is small ($ {\rm eV}<T\sim (m_{DM}^2-m_X^2)/(2m_{DM})<$ MeV), there can be resonant production of $X$ which like the case of coannihilation can satisfy the bounds. However, such fine tuned splitting between $X$ and DM is theoretically difficult to explain, especially  that since they couple together to neutrinos, one should be boson and the other should be  a fermion. Taking $m_\phi-m_{DM}\sim m_{DM}/20$  and $E_{res}\sim 100$ keV,   we find $m_{DM}\sim {\rm few} $ MeV. Larger $m_{DM}$ and/or smaller $E_{res}$ require higher degree of fine-tuning between $m_\phi$ and $m_{DM}$. Moreover for couplings large enough for efficient conversion, DM pair annihilation at $T\sim$ MeV can produce and thermalize $f$ particles before neutrino decoupling era. The produced $f \bar{f}$ will contribute to effective relativistic degrees of freedom on which there are strong bounds.

Because of issues enumerated above, we shall not try to build a model to embed the resonant scattering off DM scenario. It is however instructive to discuss
the back reaction for this scenario:
 {\it i.e.,} $f+{\rm DM}\to \nu+{\rm DM}$.
 During  the period
$m_\nu \ll T<m_{DM}$, the masses of $\nu$ and $f$ as well as the recoil energy of DM can be neglected: $E_\nu\simeq |\vec{p}_\nu| \simeq E_f\simeq
|\vec{p}_f|$. Moreover, for $s$-wave interactions, the spin of $f$ and $\nu$ have to be the same, too. As a result, the cross section of scattering and back scattering will be equal $\sigma(\nu+{\rm DM}\to f+{\rm DM})=\sigma(f+{\rm DM}\to\nu+{\rm DM})$.
As a result, the mean free path of $\nu$ and $f$ will be equal. Moreover, these interactions do not change the number density of DM.
 If the number of scatterings that a neutrino undergoes is $k$, its contribution to $\nu$ and $f$ population will be respectively equal to  $(1+(-1)^k)/2$ and  $(1-(-1)^k)/2$.
Suppose during a certain period of time (macroscopically large time scale but much smaller than $H^{-1}$),  the average number of interactions that a neutrino undergoes is $\lambda$. The distribution of number of scattering $k$ will be given by Poisson distribution so the average probability of neutrinos  not to be converted will be
$$ \frac{1}{2} < \bar{p}= \sum_{k=0}^\infty \frac{\lambda^k e^{-k}}{k !} \frac{1+(-1)^k}{2}=\frac{1+e^{-2\lambda}}{2}\leq 1.$$
For $\lambda\gg 1$, $\bar{p}$ will quickly converge to $1/2$. This is the limit that $f$ reaches thermodynamical equilibrium with $\nu$ and as a result, the entropy of neutrinos will  be shared with $f$ . Since they are both fermions, their share of entropy will be equal so $\bar{p}=1/2$ is expected from a thermodynamical perspective, too.
 With this mechanism, it will not be possible to completely remove neutrinos.

 Back reaction of neutrinos in the case of coannihilation ($\nu+ \stackrel{(-)}{\nu}\to   f +\stackrel{(-)}{f}$) is more complicated. When the temperature just approaches to the resonance  ($T \to m_X$), the density of final states is still low so back reaction is negligible. Eventually when a significant fraction of neutrinos convert into $f$ particles,  their density will become large enough to make the back reaction efficient. Let us take $R$
to be the rate of scattering of $\nu$ off any of neutrinos in the ensemble. Three regimes can be distinguished: 1) If $\int R~ dt\ll 1$, the back reaction can be neglected. 2)  If $\int R~ dt\sim 1$, the back reaction is important
but thermodynamical equilibrium has not been reached yet; 3)  If $\int R~ dt\gg 1$,
the reaction and back reaction rates will become equal. Obviously, in neither of these cases, it is possible to completely remove neutrinos.  We will focus on the third possibility in this paper. 
Notice that the energies of initial and final states in $\nu+\stackrel{(-)}{\nu}\to   f +\stackrel{(-)}{f}$ are the same. Since each neutrino in the medium undergoes reaction, we expect the energy distribution of $\stackrel{(-)}{f}$ particles to be similar to those of $\stackrel{(-)}{\nu}$. $T$-reversal symmetry implies that $\sigma
(\nu +\stackrel{(-)}{\nu}\to f +\stackrel{(-)}{f})= \sigma
(f +\stackrel{(-)}{f}
\to \nu +\stackrel{(-)}{\nu})$. The equality of reaction and back reaction rates therefore implies that the number density of $\nu$ and $f$ should be equal.

In all of the above cases, it is possible to further suppress the final density of neutrinos by converting the produced $f$ (or $\bar{f}$) to other new states that do not interact with neutrinos. Intuitively, this can be understood the following way: If $f$ particles are eliminated before they find enough time to reproduce active neutrinos, conversion of neutrinos will be more efficient. Elimination of new states can proceed via a number of processes; {\it e.g.,} $f$ particles can oscillate to new particles or they can decay into new particles. However, the above argument about back reaction applies here, too. If the process is fast enough to remove significant fraction of $f$ particles, the back reaction will be efficient in reproducing them. If all these processes come to equilibrium, the final density of neutrinos will be reduced by a factor of
\be
\label{reduction}
\frac{\rho_{\rm massive, final}}{\rho_{\rm massive, initial}} = \frac{3}{3+N}
\ee
where  $N$ is  the number of degrees of freedom that come to equilibrium with neutrinos below $T\sim 1$ MeV. Similar relation holds valid for the case that neutrino and antineutrino directly produce all these final states. For simplicity, we shall employ this last option to increase $N$.

In summary, we discussed the possibility of converting active neutrinos to lighter new particles through  resonant neutrino (antineutrino) pair coannihilation or resonant scattering of neutrinos off the dark matter particles. In case of coannihilation, this requires the intermediate state to have a mass in the range of 100~eV-100~keV. For neutrino scattering off DM, the intermediate state has to be quasi-degenerate with DM with a splitting of 100 eV-100 keV. In either case, the back scattering will be non-negligible. In the limit that reaction and back reaction come to equilibrium,  the number density of ordinary  active neutrino at recombination will be suppressed by a factor shown in Eq. (\ref{reduction}). There will be however no significant contribution to extra relativistic degrees of freedom. That is   $N_{\rm eff}$ will remain equal to the value predicted in the SM ({\it i.e.,} $N_{\rm eff}=3.046$). The contribution of ordinary neutrinos to
$N_{\rm eff}$ below resonance temperature will be given by 
\be \label{Nmass} N_{\rm massive}=\frac{3}{1+N/3}\ . \ee
The rest ($N_s=N_{\rm eff}-N_{\rm massive}$) will be in the form of lighter new particles.

\section{Models for the coannihilation scenarios\label{model}}

 As we tentatively discussed in the previous section and shall quantify more systematically in the next section, more than one new particle may be needed to make $m_\nu\sim 0.2-2.2$ eV  compatible with cosmological bounds. In our models, we assume that all these new particles are produced in the decay of resonant states that are in turn  produced by neutrino  co-annihilation:
 $$\nu +\stackrel{(-)}{\nu}\to X^*\to f_i +\stackrel{(-)}{f_i}$$
 For simplicity, we drop the index $i$.
 From model building point of view, increasing the number of final species is straightforward.

In this subsection, we first introduce a model for neutrino pair coannihilation via a new gauge interaction. We then introduce a Majoron model. In the end, we discuss various observational bounds.

{\it New $U(1)^\prime$ gauge interaction}: The active neutrinos ($\nu_L$) as well as the new sterile neutrinos ($\nu_s$) may have an interaction term of the following form with the new gauge boson:
\be \label{gauge-terms}
g^\prime (e_a^\prime \bar{\nu}_L \gamma^\mu \nu_L+e_s^\prime \bar{\nu}_s \gamma^\mu \nu_s)Z^\prime_\mu \ .\ee
The interaction leads to an $s$-channel annihilation $\nu(k_1)\bar{\nu}(k_2) \to Z^{\prime *} \to \nu_s(p_1) \bar{\nu}_s(p_2)$ with amplitude square given by
$$|M|^2=\frac{8g^{\prime 4} (e_a^\prime e_s^\prime)^2( k_1\cdot p_1 k_2\cdot p_2+k_1\cdot p_2 p_1\cdot k_2 ) }{(s-m_{Z^\prime}^2)^2+\Gamma_{Z^\prime}^2m_{Z^\prime}^2}$$
in which $s$ is the Mandelstam variable and $\Gamma_{Z^\prime}$ is  the decay width of $Z^\prime$. The cross section in the center of mass frame will be given by
\be \label{g-int}\frac{d\sigma}{d\cos \theta_a}=\frac{ g^{\prime 4} (e_a^\prime e_s^\prime p_1)^2(1+\cos^2 \theta_a)}{8\pi v_{rel}\left( (4p_1^2-m_{Z^\prime}^2)^2+\Gamma_{Z^\prime}^2m_{Z^\prime}^2\right)}\ee
where $\theta_a$ is the angle between $\vec{p}_1$ and $\vec{k}_1$. Using the above formula and the formula for conversion rate ($R$) derived in the appendix, we find that the condition $R H^{-1}>1$ implies
\be \label{value} g^\prime e_a^\prime >5 \times 10^{-11} (\frac{m_{Z^\prime}}{\rm keV})^{1/2}\ . \ee
 Since ordinary active neutrinos form a doublet along with the left-handed charged fermions, we in general expect the corresponding charged lepton to be charged under $U(1)^\prime$, too.
 In particular, if $\nu_e$ couples to $Z^\prime$, we expect the electron to couple to $Z^\prime$, too. There are strong upper bounds ($\sim 10^{-13}$) on the coupling of the electron to $Z^\prime$ of mass
(keV) from stellar coupling consideration \cite{stella} which are two orders of magnitude stronger than the values of $g^\prime e_a^\prime$ required for successful active-sterile conversion (see Eq. (\ref{value})).
There are two ways to avoid this strong constraint: (1) Remember that in the SM, photon, being a special linear combination of $W_\mu^3$ and $B_\mu$, couples to charged leptons but not to the neutrinos.
One can in principle invoke a similar mechanism by mixing the $U(1)^\prime$ gauge boson and the neutral component of the $SU(2)$ gauge bosons through the vacuum expectation value of a scalar doublet
charged under $U(1)^\prime$ to prevent the coupling of $Z^\prime$ to charged leptons while $e_a^\prime\ne 0$. We will not however elaborate further on this possibility, here. (2) We can assume that the first generation of fermions are neutral under $U(1)^\prime$ and do not couple to $Z^\prime$. As a result, the bound from stellar cooling will be automatically avoided because stars contain only first generation fermions.  This possibility has been entertained in various anomaly free $L_\mu-L_\tau$ model as well as the model presented in \cite{me}. During $T\sim$ keV, the time required for oscillation of $\nu_e$ into $\nu_{\mu,\tau}$ is much shorter than the  Hubble
time ({\it i.e.,} $ \Delta m_{21}^2/T \gg H$). As a result, while $\nu_\mu$ and $\nu_\tau$ convert to $\nu_s$, the electron neutrino in the medium will also oscillate into $\nu_\mu$ and $\nu_\tau$
and they will all come to equilibrium. In other words, because of the fast oscillation, the absence of coupling of $\nu_e$ to $Z^\prime$ will not change the picture.
A simple way to avoid anomalies is to 
take $U(1)^\prime=L_\mu-L_\tau$ for the SM fermions and to assign opposite  $U(1)^\prime$ charges to the pairs of $\nu_{si}$.

 The mass of $Z^\prime$ can come either from St\" uckelberg mechanism or from a new scalar ($\phi$) singlet under $SU(3)\times SU(2)\times U(1)$ but charged under $U(1)^\prime$ with $\langle \phi \rangle\sim m_{Z^\prime}/(g^\prime e_\phi^\prime )\sim 10~{\rm TeV}(e_a^\prime/e_\phi^\prime)$.

{\it Majoron interaction}:
Let us now consider another scenario which converts active neutrinos to lighter sterile neutrinos  through resonant production of an intermediate scalar  $J$ of keV mass. The effective couplings can be written as
$$ (\frac{g_{a}}{2}\nu_a^T c\nu_a +\frac{g_{s}}{2} \nu_{s }^T c\nu_{s }) J\ ,$$ where $c$ is a $2\times 2$ antisymmetric matrix with off-diagonal elements equal to $\pm 1$ acting on spinorial indices.
When the temperature reaches $O(m_{J})$, we can have resonant production of $J$ and its subsequent decay. At the center of mass frame,
\be \label{MajAmp} 
|M|^2=\frac{ g_a^2 g_s^2 (2 k_1\cdot k_2) (2 p_1\cdot p_2)}{(s-m_J^2)^2+m_J^2\Gamma_J^2}\ee where $\Gamma_{J}$ is the total decay rate of the intermediate scalar. 
The cross section is therefore given by
$$\sigma (\nu_a(k_1)+\nu_a(k_2) \to J^* \to \nu_s(p_1) +\nu_s(p_2)) \sim \frac{g_a^2g_s^2}{4\pi} \frac{p_1^2}{(s-m_{J}^2)^2+\Gamma_{J}^2m_{J}^2}$$
Similarly to Eq. (\ref{value}) for efficient conversion of active neutrinos to sterile
ones, $g_a$ should satisfy the following bound $$g_a> 5 \times 10^{-11}(m_J/{\rm keV})^{1/2}.$$

Let us now consider the high energy completion of the model. The effective $g_a$ coupling may come from mixing with  a heavy $SU(2)$ triplet $\Delta$:
$$\frac{g_\Delta}{2} L^Tc\epsilon \Delta L +\frac{g_{s}}{2} J \nu_s^T c \nu_s$$
where $\epsilon$, like $c$, is a $2\times 2$ antisymmetric matrix  with off-diagonal elements equal to $\pm 1$, but unlike $c$, acts on the electroweak $SU(2)$ indices.
The masses of scalars are given by
\be \label{scalarMs} \frac{m_{J}^2}{2} J^2+m_\Delta^2 {\rm Tr}[ \Delta^\dagger \Delta]+ \frac{\lambda_{\Delta J}}{2} H^\dagger \Delta\epsilon H^* J.\ee
Of course $m_\Delta$ should be larger than electroweak scale; otherwise, the components of $\Delta$ would have been discovered by now at colliders.
The mixing is given by
$\alpha \simeq \lambda_{\Delta J} \frac{v^2}{m_\Delta^2-m_J^2}$ and $g_a\simeq g_\Delta \alpha$.
Taking $g_a \sim 5 \times 10^{-11}$, $g_\Delta>0.1$ and $m_\Delta \sim 1$ TeV, we find that $\alpha^2 m_\Delta^2\ll {\rm keV}^2$ so the $\lambda_{\Delta H}$ term does not considerably change the mass eigenvalues.  Taking $m_J^2$ in Eq. (\ref{scalarMs})
 to be of order of $({\rm keV})^2$, we will naturally obtain $J$ mass equal to keV without any fine tuning despite the large hierarchy between $m_\Delta$ and $m_J$.
The production rate of $J$ via $\lambda_{\Delta J}$ coupling  at high temperatures is given by $\lambda_{\Delta J}^2 T/(4\pi)\sim  T(g_a^2/g_\Delta^2 )(m_\Delta^2/v^2)^2 /(4\pi) $ which for $g_a\sim 5 \times 10^{-11}$ will be much smaller than $H|_{T=m_\Delta}$. Thus, no extra contribution to relativistic degrees of freedom at big bang nucleosynthesis era is predicted.

{\it More observational bounds:} Several observational bounds have been already discussed above. Let us now review other potential bounds. The required values of new coupling of active neutrinos within this scenario are so small that they can easily avoid all existing bounds. Bounds from supernova cooling consideration are of  order of $10^{-7} $ \cite{ancient} which are four orders of magnitude weaker than the required value for coupling (see Eq. (\ref{value})). The bounds from terrestrial experiment (rare meson decay) are even weaker \cite{Lessa}

As shown in \cite{secret}, for the case of massless Majoron, very strong bounds can be obtained from the free streaming of neutrinos at recombination era $T\sim 0.3$ eV.
In our case, we have to also make sure that  active neutrinos as well as the final particles that have been produced during $T\sim m_X$ freely stream at recombination.
Since we have taken the couplings of $\nu_s$ to be larger than that of active neutrinos, it is enough to
 check if $\nu_s$ stream freely during recombination ($T\sim $ 0.3 eV).  Let us first consider the Majoron interaction:
$\sigma_{scattering}\sim \frac{g_s^4 T_\nu^2}{4 \pi m_J^4}$ and $\Delta t\sim 350000 ({\rm 0.3 ~eV}/T)^2 ~{years}$ therefore $$n_\nu \sigma_{scattering} \Delta t|_{T\sim 0.3~{\rm eV}}\sim  10^{9} g_s^4 \left( \frac{\rm keV}{m_{J}}\right)^4.$$
Thus, for $g_s \lesssim 0.005(m_J/{\rm keV})$, the sterile neutrinos  freely stream.
For gauge interactions, the $g_s/m_J$ ratio has to be just replaced by $g^\prime e^\prime_s/m_{Z^\prime}$.
The reason why strong bounds found in \cite{secret} do not apply here is that while in \cite{secret} the Majoron is taken to be massless or very light, in our case $m_J\gg 0.3$ eV.


\section{Cosmological constraints\label{Cosmo}}

The minuteness of the coupling between neutrinos and the new scalar has two important implications: First, it means that the sterile neutrinos are not thermalized  prior to the decoupling of active neutrinos. Second, the light scalars are never thermalised.

Once the active and sterile neutrinos equilibrate at $T \sim m_X \sim $ keV the total energy density in neutrinos, sterile neutrinos and scalars is fixed at the standard model value $N_{\rm eff} = 3.046$. Because of the resonant nature of the production all neutrinos are highly relativistic at the time of production and there is no additional contribution to $N_{\rm eff}$ from rest mass effects (unlike for example the neutrinoless universe scenario \cite{Beacom:2004yd,Hannestad:2004qu,Archidiacono:2014nda,Archidiacono:2015oma}).

Furthermore, below the resonance temperature the interaction remains unimportant because the coupling is so small and the mass of the scalar is high. This means neutrinos and sterile neutrinos remain weakly interacting and that both neutrinos and sterile neutrinos free stream like ordinary neutrinos.

From the point of view of CMB and structure formation the scenario is therefore the following: The total relativistic energy density in neutrinos and sterile neutrinos is given by $N_{\rm eff} = N_s + N_{\rm massive}$, where $N_{\rm massive}$ gives the energy density remaining in the massive standard model neutrinos and $N_s$ gives the energy density in the massless sterile neutrino component.

We have performed a likelihood analysis of current data using \texttt{CosmoMC} \cite{Lewis:2002ah}.
Our benchmark CMB data set consists of the Planck 2015 high multipole temperature data and low multipole polarization data (PlanckTT+lowP), implemented according to the prescription of Ref. \cite{Aghanim:2015xee}.
We have also performed the analysis with
Baryonic Acoustic Oscillation (BAO) data,including 6dFGS \cite{Beutler:2011hx}, SDSS-MGS \cite{Ross:2014qpa}, BOSS-LOWZ BAO \cite{Anderson:2012sa} and CMASS-DR11 \cite{Anderson:2013zyy}.
The neutrino sector is described by the parameters $N_{\rm massive}$ and the physical neutrino mass $m_\nu$.

The other cosmological parameters used in the analysis correspond to those in the standard Planck 2015 analysis of neutrino mass:
The baryon density, $\Omega_b h^2$, the cold dark matter density, $\Omega_c h^2$, the angular scale of the first CMB peak, $\theta$, the optical depth to reionization, $\tau$, the amplitude of scalar fluctuations, $A_s$, and the scalar spectral index, $n_s$.

\begin{figure}
\begin{center}
\includegraphics[width=0.4\columnwidth]{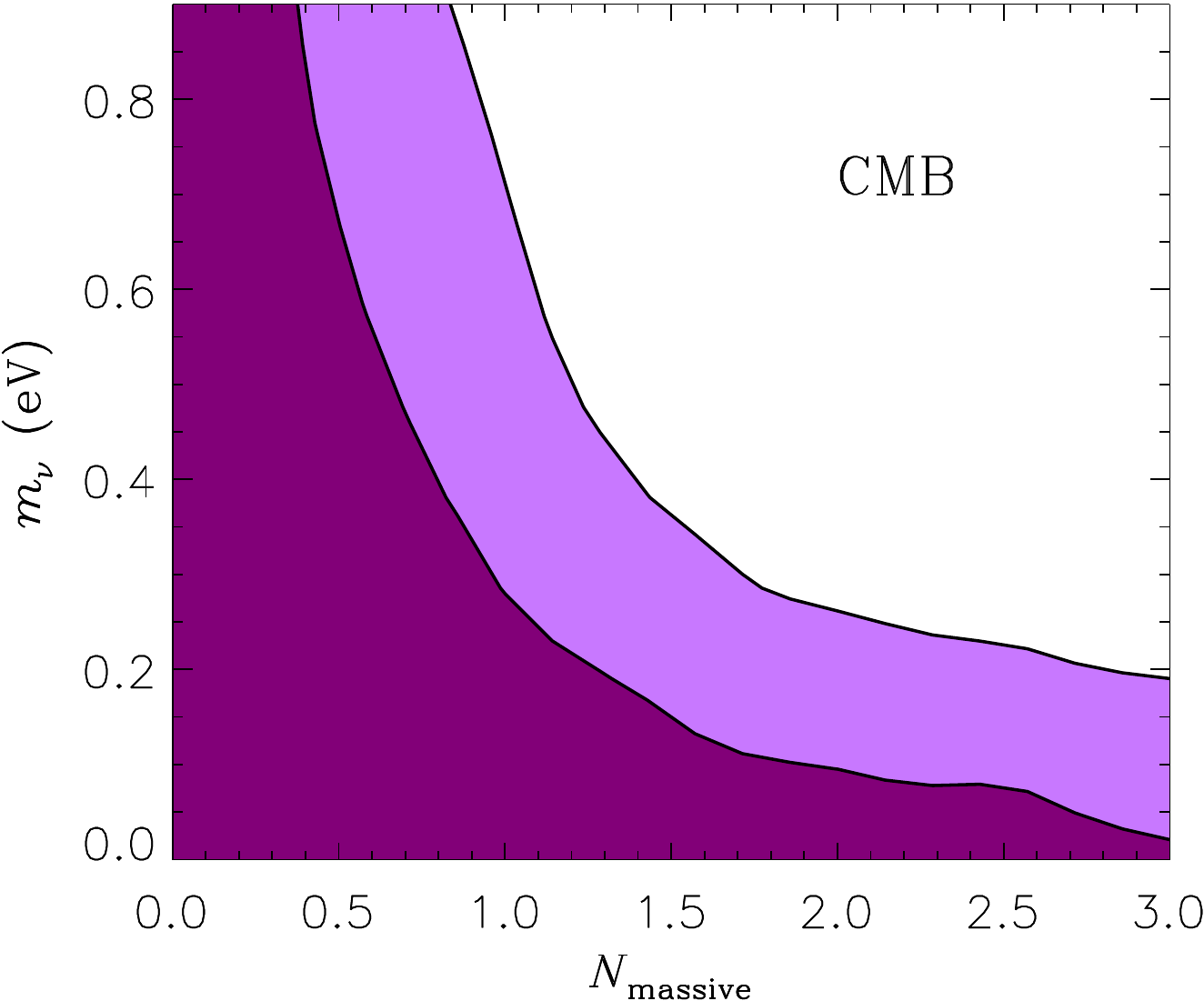}\includegraphics[width=0.4\columnwidth]{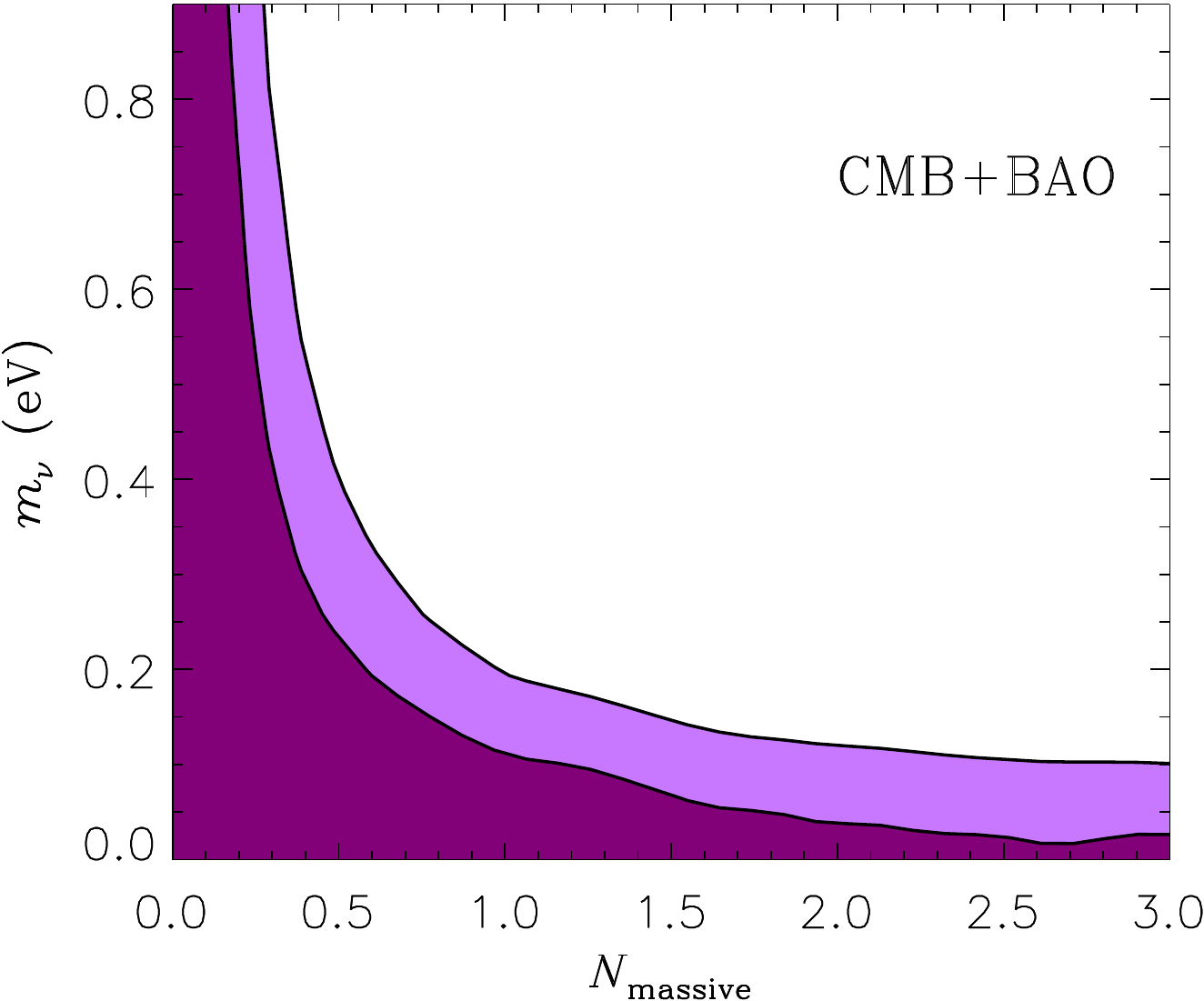}
\end{center}
\caption{2D marginalized 68\% and 95\% likelihood contours for the parameters $N_{\rm massive}$ and $m_\nu$. The left panel shows CMB data only and the right panel includes BAO data.
}
\label{fig:likelihood}
\end{figure}

The results are shown in Fig.~\ref{fig:likelihood}. With the inclusion of CMB data only three massive neutrinos of degenerate mass 0.2 eV are never disfavored at more than 95\% C.L. (fitting well with the formal Planck 2015 bound from CMB data of $\sum m_\nu < 0.72$ eV at 95\% C.L.), and if $N_{\rm massive}$ is suppressed to 1.2 the scenario is compatible with observations at the 68\% C.L.

Once BAO data is included massive neutrinos are disfavored at much higher significance, again fitting well with the Planck 2015 bound of $\sum m_\nu < 0.21$ eV for the standard model case. For the case of $N_{\rm massive}=3.046$ (the standard model case) a single neutrino mass of 0.2 eV is disfavored at close to $5\sigma$.
However, provided that $N_{\rm massive}$ is shifted to down approximately 1 the model is only disfavored at 95\% C.L. (as could be expected because it corresponds to a single mass state of $m_\nu \sim 0.2$ eV), and if $N_{\rm massive} \sim 0.6$ the model shifts to the 68\% region.

Our expectation that the presence of resonant conversions makes massive neutrinos more compatible with cosmological data is therefore confirmed and could indeed be a possible explanation if KATRIN measures a mass for the standard model neutrinos. {
In particular, $m_\nu=0.35$ eV which is the discovery limit of KATRIN at 5 $\sigma$ \cite{0.35} can be made compatible with the 95 \% C.L. limits from CMB (CMB+BAO)  provided that $N_{massive}$ is lowered down to 1.5 (0.6) which according to Eq. (\ref{Nmass}) can be achieved if for each active flavors, there are 1 (4) light or massless sterile neutrinos that are  produced in the resonance ({\it i.e.,}   $N=3~(12)$).}

\section{Summary and concluding remarks\label{Conclusions}}
To  relax cosmological bounds on the neutrino mass we have introduced scenarios within which neutrinos are converted to lighter particles in the era after neutrino decoupling from SM particles and before recombination. Since the conversion takes place after neutrino decoupling, $N_{\rm eff}$ remains equal to 3.046 as in the SM.
The energy distribution of the final particles is similar to that of neutrinos. The conversion of neutrinos to the new states and the inverse process can equilibrate the new species so that the contribution of active neutrinos to $N_{\rm eff}$ will be suppressed by a factor of $3/(3+N)$ where $N$ is the number of final light stable states that neutrinos convert into.


{We have found that if only CMB measurements are used, there is no significant need for dilution of the massive neutrino states in order to remain within the 95\% C.L. limits. This confirms the results of \cite{Ade:2015xua}. To make $m_\nu=0.2$~eV compatible with the 68 \% C.L. limit of (only)  CMB data,  the requirement is that $N \gtrsim 3.5$. Adding Baryon Acoustic Oscillation considerations more dilution will be required: We need $N \gtrsim 6$ in order  for the model with $m_\nu=0.2$~eV to remain compatible within the 95\% C.L. bounds with current CMB+BAO measurements. For neutrino masses larger than the  KATRIN limit of 0.2 eV, the number of additional states must be correspondingly larger.}

The scenario has to satisfy the following three requirements simultaneously: 1) efficient conversion after neutrino decoupling; 2) negligible production before neutrino decoupling and 3)  free streaming of neutrino and the new stable light states during recombination. We have shown that  the resonant interaction of neutrinos at eV$\ll T\ll$ MeV can satisfy all these three requirements.  We have introduced two classes of possible conversion scenarios: 1) scattering of active neutrinos off dark matter. To make the resonant conversion successful, there should be a new particle quasi-degenerate with dark matter with splitting of $O(keV)$ and spin difference of 1/2. 2) Resonant annihilation of neutrino and/or antineutrino pair by production of an intermediate state of mass keV which immediately decays to new lighter states. For successful conversion, the coupling of neutrinos to the intermediate state should be larger than $5 \times 10^{-11}$.

We have introduced two specific models to realize the second scenario: In the {\it first model} the intermediate state is a gauge boson of mass keV which couples to the second and third generations of leptons but not to the first generation to avoid the stringent bounds from star cooling.  The gauge symmetry  ($U(1)^\prime$) in question can be for example $L_\mu-L_\tau$. To maintain anomaly cancelation the new particles can be scalars or vector-like fermions (or equivalently pairs of Weyl fermions with opposite $U(1)^\prime$ charges).  In the {\it second model}, the intermediate state is a scalar of keV mass with a Majoron type coupling to neutrinos.
We have shown that the model can be naturally UV-completed by introducing a heavy $SU(2)$ triplet scalar mixed with a light singlet.

Finally, we again wish to stress that a detection of a non-zero mass for the active neutrinos by KATRIN will be extraordinarily interesting because new physics must be invoked to make the measurement compatible with cosmology.

\section{Appendix}
In the following, we calculate the rate of interaction of a neutrino of four-momentum $P_2^\mu=(p_2,0,0,p_2)$ with any other neutrino in medium at temperature
$T$:
$$R=\frac{1}{2p_2} \int\int\int \frac{d^3 p_1}{(2\pi)^32 p_1}\frac{1}{1+\exp^{p_1/T}} \frac{d^3 k_1}{(2\pi)^32 k_1}\frac{d^3 k_2}{(2\pi)^3 2k_2} (2\pi)^4 \delta^4(p_1+p_2-k_1-k_2) |M|^2$$
where $|M|^2$ close to resonance is given by Breit-Wigner function as
$$|M|^2=\frac{\mathcal{A}f(\theta_{a})}{(q^2-m_{X}^2)^2+ m_{X}^2 \Gamma_{X}^2},$$
where $\mathcal{A}$ is almost constant and $q=p_1+p_2$ is the four-momentum of the intermediate boson. $\theta_{a}$ is the angle between momenta of initial $\nu_a$ and final particle in the center of mass frame.
From Eq. (\ref{g-int}), we read that in the case of gauge interactions ($X=Z^\prime$)
$$\mathcal{A}=g^{\prime 4} (e^\prime_a e^\prime_s)^2m_{Z^\prime}^4\ \ {\rm and} \ \ f(\theta_{a})=1+\cos^2\theta_a .$$
 From Eq. (\ref{MajAmp}), we read that in case of Majorana interactions
$$\mathcal{A}=g_a^2g_s^2 (2 k_1\cdot k_2)(2 p_1\cdot p_2)|_{resonance}=g_a^2g_s^2m_J^4 \ \ {\rm and} \ \ f(\theta_{a})=1.$$
Using narrow width approximation we find $$|M|^2\simeq  B f(\theta_a)\delta(q^2-m_{X}^2) ~~~~{\rm where}~~~~ B\equiv\frac{\mathcal{A} \pi}{m_{X}\Gamma_{X}}.$$
Remember that $\Gamma_X$ is the total decay width. If new particles dominate the decay of $X$ ({\it i.t.,} if $e^\prime_s\gg e^\prime_a$ for gauge interactions or if $g_s\gg g_a$ for Yukawa interactions), $B$ will be independent of the couplings to new states and will be given by the coupling to active neutrinos.  That is for gauge interactions, $$ \Gamma_X\simeq \frac{N(g^{\prime } e^{\prime }_s)^2 m_{Z^\prime}}{8 \pi} \ \ \ {\rm so} \ \ \ B=\frac{8\pi^2 (g^\prime e_a^\prime m_{Z^\prime})^2}{N}$$ and for Yukawa interaction
 $$ \Gamma_X\simeq \frac{N g_s^2 m_{J}}{16 \pi} \ \ \ {\rm so} \ \ \ B=\frac{16\pi^2 g_a^2 m_{J}^2}{N}.$$
 In both cases, $N$ is the number of new states coupled to $X$.
We can simplify the calculation by the following convolution:$$
R= \frac{(2\pi)^4 B}{(2\pi)^92p_2} \int \frac{d^3 p_1}{2p_1} \frac{1}{1+\exp^{p_1/T}}\int d^4 q \delta^4(p_1+p_2-q)S(q^2)$$
where
$$S(q^2)\equiv \int \frac{d^3 k_1}{2k_1}\int \frac{d^3 k_2}{2k_2} \delta^4(k_1+k_2-q)f(\theta_a) \delta(q^2- m_X^2)=\frac{ \pi}{2}b \delta(q^2-m_X^2)\ ,$$
where $b=1 (4/3)$ for Majorana (gauge) interaction.
To calculate $S(q^2)$, we  have used its Lorentz invariance and have performed  calculation in the rest frame of intermediate $X$.
Remembering that $q^2=p_1p_2(1-\cos \theta)$ (in which $\theta$ is the angle between initial momenta), we can  write
$$R=\frac{B\cdot b}{2^8\pi^3 p_2}\int \frac{p_1^2 dp_1d\cos\theta}{p_1}\frac{1}{1+\exp^{p_1/T}}{\delta( p_1p_2(1-\cos\theta)-m_X^2)}=$$
$$\frac{B\cdot b \cdot T}{2^8\pi^3 p_2^2}\left(\log(1+\exp^{m_X^2/2p_2T})-m_X^2/2p_2T\right).$$
For $p_2T\ll m_X^2$, we can write $R \to B\cdot b\cdot T/(2^8 \pi^3 p_2^2)\exp^{-m_X^2/2p_2T}$. For $p_2 T\gg m_X^2$, we can write $R\to \log(2)B\cdot b \cdot T/(2^8\pi^3 p_2^2)$.
 To check if the active-sterile conversion is effective,
$R$ should be compared to $H=T^2/M_{Pl}^*$. The majority of neutrinos have energy of order of temperature $p_2
\sim T$ so for high temperatures $T\gg m_X$, the conversion of active neutrinos  (for majority of neutrinos in the medium) to sterile neutrinos are not effective. In other words, for $T\gg m_X$ and $ p_2\sim T$, we expect $R/H\ll 1$. (Notice that even at high temperatures if $p_2$ is sufficiently small, their conversion rate
to sterile neutrinos will be relatively high but such low energy neutrinos comprise only small fraction of neutrinos.)
Moreover, at low temperatures for which $p_2T\ll m_X^2$, $R/H$ is also small so conversion will be negligible.
Conversion can be efficient ($R \gtrsim H$)  only at $T\sim m_X$ provided that $R|_{T\sim m_X}\sim H|_{T\sim m_X}$. Taking $m_X\sim 1$~keV, this condition can be translated into
\be  g_a \ {\rm or} \ g^\prime e^\prime_a \gtrsim 5\times 10^{-11}.\ee Remember that we have assumed that the coupling of $X$ to the new lighter particles are stronger.
\subsection*{Acknowledgments}
          The authors would like to thank Mainz Institute for Theoretical Physics and organizers of ``Crossroads of Neutrino Physics" extended workshop where this project started for kind and generous hospitality.
YF would like to acknowledge partial support from the  European Union FP7 ITN INVISIBLES (Marie Curie Actions, PITN- GA-2011- 289442).


\begin{thebibliography}{99}

\bibitem{Gonzalez-Garcia:2014bfa}
  M.~C.~Gonzalez-Garcia, M.~Maltoni and T.~Schwetz,
  JHEP {\bf 1411}, 052 (2014)
  [arXiv:1409.5439 [hep-ph]].



\bibitem{Mainz}
J. Bonn {\it et al.,} Nucl. Phys Proc. Suppl. \textbf{91} (2001) 273.

\bibitem{Osipowicz:2001sq}
  A.~Osipowicz {\it et al.} [KATRIN Collaboration],
  hep-ex/0109033;
  L.~Bornschein [KATRIN Collaboration],
  Proceedings of the Fifteenth Lomonosov Conference on Elementary Particle Physics
Moscow, Russia, 18 ֠24 August 2011.
  \bibitem{0.35}
  M.~Beck [KATRIN Collaboration],
  J.\ Phys.\ Conf.\ Ser.\  {\bf 203} (2010) 012097
  [arXiv:0910.4862 [nucl-ex]].
  \bibitem{Mertens}
   S.~Mertens {\it et al.},
  JCAP {\bf 1502} (2015) 02,  020
  [arXiv:1409.0920 [physics.ins-det]].

	
\bibitem{Ade:2015xua}
  P.~A.~R.~Ade {\it et al.} [Planck Collaboration],
  arXiv:1502.01589 [astro-ph.CO].
	

\bibitem{Laureijs:2011gra}
  R.~Laureijs {\it et al.} [EUCLID Collaboration],
  arXiv:1110.3193 [astro-ph.CO].
	
\bibitem{Hamann:2012fe}
  J.~Hamann, S.~Hannestad and Y.~Y.~Y.~Wong,
  JCAP {\bf 1211}, 052 (2012)
  [arXiv:1209.1043 [astro-ph.CO]].
	
\bibitem{Basse:2013zua}
  T.~Basse, O.~E.~Bjaelde, J.~Hamann, S.~Hannestad and Y.~Y.~Y.~Wong,
  JCAP {\bf 1405}, 021 (2014)
  [arXiv:1304.2321 [astro-ph.CO]].
	
\bibitem{Audren:2012vy}
  B.~Audren, J.~Lesgourgues, S.~Bird, M.~G.~Haehnelt and M.~Viel,
  JCAP {\bf 1301}, 026 (2013)
  [arXiv:1210.2194 [astro-ph.CO]].
	
	

\bibitem{secret}
S.~Hannestad and G.~Raffelt,
  Phys.\ Rev.\ D {\bf 72} (2005) 103514  [hep-ph/0509278].  
 
  \bibitem{extraPLANCK}
P.~A.~R.~Ade {\it et al.} [Planck Collaboration],
  Astron.\ Astrophys.\  {\bf 571} (2014) A16
  [arXiv:1303.5076 [astro-ph.CO]].
 \bibitem{stella}
   P.~Arias, D.~Cadamuro, M.~Goodsell, J.~Jaeckel, J.~Redondo and A.~Ringwald,
  JCAP {\bf 1206} (2012) 013
  [arXiv:1201.5902 [hep-ph]].
\bibitem{me}
Y.~Farzan,
  Phys.\ Lett.\ B {\bf 748} (2015) 311
  [arXiv:1505.06906 [hep-ph]].
  \bibitem{ancient}
  Y.~Farzan,
  Phys.\ Rev.\ D {\bf 67} (2003) 073015
  [hep-ph/0211375].
\bibitem{Lessa}
A.~P.~Lessa and O.~L.~G.~Peres,
  Phys.\ Rev.\ D {\bf 75} (2007) 094001
  [hep-ph/0701068];
{\it For an update see}  Y.~Farzan,
  Mod.\ Phys.\ Lett.\ A {\bf 25} (2010) 2111
  [arXiv:1009.1234 [hep-ph]].
\bibitem{Beacom:2004yd}
  J.~F.~Beacom, N.~F.~Bell and S.~Dodelson,
  Phys.\ Rev.\ Lett.\  {\bf 93}, 121302 (2004)
  [astro-ph/0404585].

\bibitem{Hannestad:2004qu}
  S.~Hannestad,
  JCAP {\bf 0502}, 011 (2005)
  [astro-ph/0411475].
	

\bibitem{Archidiacono:2014nda}
  M.~Archidiacono, S.~Hannestad, R.~S.~Hansen and T.~Tram,
  Phys.\ Rev.\ D {\bf 91}, no. 6, 065021 (2015)
  [arXiv:1404.5915 [astro-ph.CO]].
	
\bibitem{Archidiacono:2015oma}
  M.~Archidiacono, S.~Hannestad, R.~S.~Hansen and T.~Tram,
  arXiv:1508.02504 [astro-ph.CO].
	
\bibitem{Lewis:2002ah}
  A.~Lewis and S.~Bridle,
  Phys.\ Rev.\ D {\bf 66}, 103511 (2002)
  [astro-ph/0205436].


\bibitem{Aghanim:2015xee}
  N.~Aghanim {\it et al.} [Planck Collaboration],
  [arXiv:1507.02704 [astro-ph.CO]].

\bibitem{Beutler:2011hx}
  F.~Beutler, C.~Blake, M.~Colless, D.~H.~Jones, L.~Staveley-Smith, L.~Campbell, Q.~Parker and W.~Saunders {\it et al.},
  Mon.\ Not.\ Roy.\ Astron.\ Soc.\  {\bf 416}, 3017 (2011)
  [arXiv:1106.3366 [astro-ph.CO]].

\bibitem{Ross:2014qpa}
  A.~J.~Ross, L.~Samushia, C.~Howlett, W.~J.~Percival, A.~Burden and M.~Manera,
  arXiv:1409.3242 [astro-ph.CO].

\bibitem{Anderson:2012sa}
  L.~Anderson, E.~Aubourg, S.~Bailey, D.~Bizyaev, M.~Blanton, A.~S.~Bolton, J.~Brinkmann and J.~R.~Brownstein {\it et al.},
  Mon.\ Not.\ Roy.\ Astron.\ Soc.\  {\bf 427}, no. 4, 3435 (2013)
  [arXiv:1203.6594 [astro-ph.CO]].

\bibitem{Anderson:2013zyy}
  L.~Anderson {\it et al.}  [BOSS Collaboration],
  Mon.\ Not.\ Roy.\ Astron.\ Soc.\  {\bf 441}, 24 (2014)
  [arXiv:1312.4877 [astro-ph.CO]].
\end{thebibliography}
\end{document}